\documentclass[aps,prb,showpacs,twocolumn,amsmath,amssymb,superscriptaddress]{revtex4}

\usepackage[colorlinks,citecolor=blue,linkcolor=cyan]{hyperref}
\usepackage[usenames,dvipsnames,svgnames,table]{xcolor}

\usepackage{graphicx}
\usepackage{Jonasmacros}
\usepackage{bm}
\usepackage{mathptmx}
\usepackage{pxfonts}

\renewcommand{\section}[1]{{\par\it #1.---}\ignorespaces}

\begin{document}
\date{\today}
\title{Engineered near-perfect backscattering on the surface of a topological insulator with nonmagnetic impurities}
\author{J. Fransson}
\email{Jonas.Fransson@physics.uu.se}
\affiliation{Department of Physics and Astronomy, Uppsala University, Box 530, SE-751 21\ \ Uppsala}

\author{A. M. Black-Schaffer}
\affiliation{Department of Physics and Astronomy, Uppsala University, Box 530, SE-751 21\ \ Uppsala}

\author{A. V. Balatsky}
\affiliation{NORDITA, Roslagstullsbacken 23, SE-106 91\ \ STOCKHOLM, Sweden}
\affiliation{Institute for Materials Science, Los Alamos, NM 87545, United States}

\begin{abstract}
We show how to engineer enhanced skew scattering and gap-like opening in the surface state of three-dimensional topological insulators using only non-magnetic impurities. Enhanced skew scattering off non-magnetic impurities is obtained as a finite size effect of the scattering potential. Intimately related to the generated skew-scattering is the emergence of a gap-like density of electron states locally around the impurities and surrounded by sharp resonances, with an extended energy gap appearing in engineered impurity structures.
\end{abstract}
\pacs{72.10.Fk, 73.20.At, 73.20.Hb, 73.90.+f}
\maketitle

Topological insulators (TIs) belong to a new state of matter, where the bulk is insulating but there are conducting surface states with properties governed by the topology of the bulk band structure.\cite{Kane_PRL05,Bernevig_2006, Kane_RMP2010, Qi11RMP} The surface states have a linear Dirac-like energy dispersion, where the electron momentum is locked to its spin.\cite{Kane_PRL07, Moore_PRB07, Qi08, Hsieh09spin, DMreview} A direct consequence of this spin-helical Dirac structure is that full $180^\circ$ back-scattering, $\bfk\rightarrow-\bfk$, requires a spin-flip, and thus conventional wisdom dictates that only magnetic impurities can give rise to such back-scattering. This is intimately related to the fact that the TI Dirac surface state spectrum can only be gapped by impurities or other perturbations breaking time-reversal symmetry.\cite{Kane_PRL07, Moore_PRB07, Qi08}

Based on this theoretical reasoning, one would expect a very clear difference in the behavior of magnetic and non-magnetic impurities on the surface of a three-dimensional (3D) TI. However, despite multiple experimental studies, using predominantly angle-resolved photoemission spectroscopy (ARPES) and scanning tunneling microscopy/spectroscopy (STM/STS), no clear consensus has appeared as to the properties of the TI surface state in the presence of impurities. 
While the surface state for magnetic impurities in the bulk or in a thin film has been shown to exhibit features resembling a gap,\cite{Chen10, Wray11, Xu12} no energy gap has been found for magnetic impurities deposited directly on the surface.\cite{Scholz12, Valla12, Honolka12PRL, Schlenk13PRL} Interestingly, several studies have found no significant difference in the behavior of magnetic and non-magnetic surface impurities.\cite{Bianchi11, Valla12}
Some suggested explanations for these results stem from natural variations in real TI materials; multiple bands can appear at the Fermi level with impurity doping,\cite{Wray11} surface band bending has been found to cause valence band quantization,\cite{Bianchi10, King11} and the effective magnetic field from some magnetic impurities might align as to not gap the surface,\cite{Honolka12PRL} to mention a few proposals.
Furthermore, the strong suppression of back-scattering off non-magnetic disorder reported from quasiparticle interference measurements using STS has been limited to multiband surfaces,\cite{Yazdani_Nature09} the high-energy regime,\cite{Xue_TI_QPI_PRL_2009,KapitulnikPRL,Yazdani_NPhys2011} where hexagonal warping is present,\cite{Fu09warping}  and long-wavelength alloy disordering.\cite{Kim14}
In addition, STS measurements have found strong impurity resonances significantly modifying the local density of states even for non-magnetic impurities \cite{Teague12, Alpichshev_PRL12} as well as step edges.\cite{Alpichshev11}

Recent experiments thus show on a clear complexity of the TI surface state in real materials. However, there is still a fundamental flaw in the argument for the insensitivity to local non-magnetic impurities even for the simplest single TI Dirac surface state. Even if $180^\circ$ back-scattering is forbidden for a single scattering process, the 2D Dirac cone structure still allows for skew-scattering. Thus, multiple scattering events, even in a fully time-reversal invariant system, can add up and effectively produce near-perfect back-scattering.

In this work we show that skew-scattering off non-magnetic impurities on the surface of a 3D TI is enhanced for spatially extended impurities or clusters of two or more point-like impurities. While we verify that multiple scattering off non-magnetic impurities cannot generate perfect $180^\circ$ back-scattering, $\bfk\rightarrow-\bfk$, we demonstrate that near-perfect back-scattering ($180^\circ\pm\delta$, $\delta>0$ infinitesimal) is always present and often prominent.
We further show that this results in an effective energy gap emerging locally in the spectrum of the surface states around the impurities. The gap-like regime is surrounded by sharp resonances, with energies depending on the spatial shape and potential strength of the impurity. 
By engineering impurity structures, an extended effectively gapped region can be produced on the surface of a 3D TI using only non-magnetic impurities.
We thus propose two closely positioned point-like non-magnetic defects as a minimal model on the 2D TI surface state in which both effective back-scattering and gap opening appear even without breaking time-reversal symmetry. Since any realistic non-magnetic defect has a continuous scattering potential, effective back-scattering is thus generally present for all realistic non-magnetic defects. These results also clearly demonstrate that realistic non-magnetic impurities can behave very similarly to what is expected from their magnetic counterparts. This offers an important piece of theoretical background for understanding the current experimental status.

To be specific, we consider the surface states of a topological insulator (TI) by means of a continuum model:
\begin{align}
\Hamil_0=&
	v\sum_\bfk\Psi^\dagger_\bfk[\bfk\times\hat{\bf e}_3]\cdot\bfsigma\Psi_\bfk\ ,
\end{align}
where $\Psi_\bfk^\dagger=(\csdagger{\bfk\up}\  \csdagger{\bfk\down})$ ($\Psi_\bfk$) is the electron creation (annihilation) spinor at momentum $\bfk$. Here, $\bfsigma$ is the Pauli spin matrix vector. Impurity scattering on the surface is accounted for by the Hamiltonian $\Hamil_\inter=\int\Psi^\dagger(\bfr)\bfV(\bfr)\Psi(\bfr)d\bfr$, where the scattering potential matrix $\bfV(\bfr)$ in general is a spatially continuous function peaked around the impurity site. To allow for analytical calculations, the impurity potential is often approximated by Dirac delta function(s), i.e.~$\bfV(\bfr)=\sum_m\bfV_m\delta(\bfr-\bfr_m)$, such that the impurity scattering occurs at discrete points $\bfr_m$ in space with potential strengths $V_m$. Here, we use this approach to simulate the effect of spatially extended impurities. We point out that the collection of two or more scattering points in this work should be regarded as a collective unit rather than individual impurities.

\begin{figure}[t]
\begin{center}
\includegraphics[width=0.99\columnwidth]{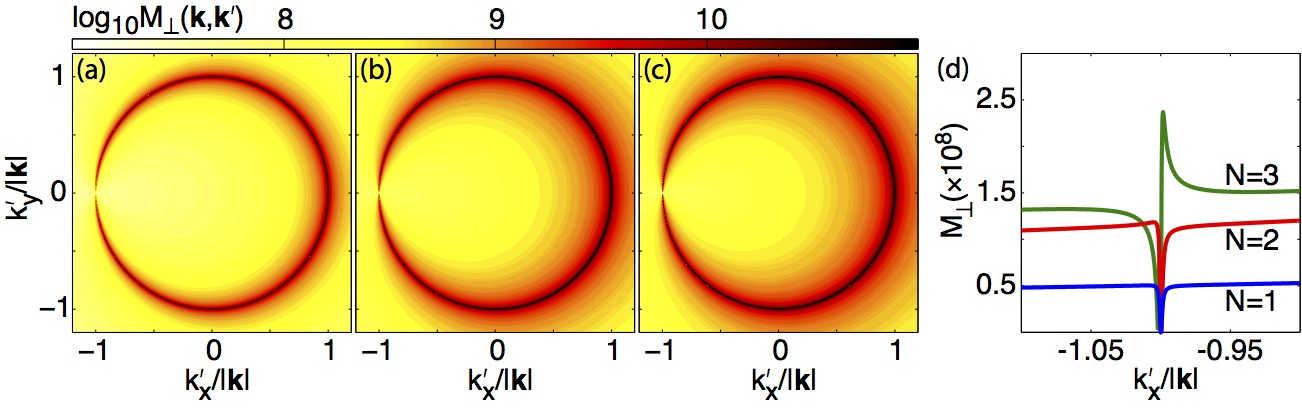}
\end{center}
\caption{Evolution of $M_\perp(\bfk,\bfk')$ for (a) one through (c) three scattering points located on a circle with 1 \AA\ radius. (d) Traces of $M_\perp(\bfk,\bfk')$ along $k_x'$ near $\bfk'=-\bfk$ corresponding to the plots in (a) | (c), where $N$ denotes the number of scattering points. Here, we have used $\bfk=k(1,0)$, $V_m=50$ meV, and cutoff $D_c=300$ meV (see e.g. Ref. \onlinecite{SCZhang09}), for $T=1$ K.
}
\label{fig-Mperp}
\end{figure}

First, for a single scattering point at $\bfr_0$ on the surface, we write the scattering potential $\bfV(\bfr)=V\sigma^0\delta(\bfr-\bfr_0)$, where $\sigma^0$ is the $2\times2$ identity matrix. Using a $T$-matrix approach, we obtain an exact and elegant description of the scattering off the defect, given by
\begin{subequations}
\label{eq-Tmatrix}
\begin{align}
\bfG(\bfk,\bfk')=&
	\delta(\bfk-\bfk')\bfg(\bfk)
	+\bfg(\bfk)e^{-i\bfk\cdot\bfr_0}\bfT e^{i\bfk'\cdot\bfr_0}\bfg(\bfk'),
\label{eq-GggTG}
\\
\bfT=&
	(1-\bfV\bfg_0)^{-1}\bfV
	=
	\sigma^0(V^{-1}-g_0)^{-1}.
\label{eq-T}
\end{align}
\end{subequations}
Here, the \emph{bare} surface Green function (GF) $\bfg(\bfk,z)=[z\sigma^0+v(\bfk\times\hat{\bf e}_3)\cdot\bfsigma]/[z^2-(vk)^2]$, $z=\omega+i\delta$, $\delta>0$ infinitesimal, while $\bfg_0(\omega)=g_0(\omega)\sigma^0=\sum_\bfk\bfg(\bfk)=\omega(2\ln|\omega|/D_c-i\pi\sign\omega)\sigma^0/4\pi v^2$, with a finite cut-off energy $D_c$ for the bandwidth of the surface states. Especially, we notice that the $T$-matrix is proportional to $\sigma^0$, which leads to the $\bfk$-dependence of the last term in Eq.~(\ref{eq-GggTG}) being governed by the product $e^{-i\bfk\cdot\bfr_0}\bfg(\bfk,z)\bfg(\bfk',z)e^{i\bfk'\cdot\bfr_0}$. The off-diagonal components of this product, causing non-forward scattering, can be written $zv(ke^{i\varphi}+k'e^{i\varphi'})/\{[z^2-(vk)^2][z^2-(vk')^2]\}$, where $\tan\varphi=k_x/k_y$. For elastic scattering we have $\bfk'=ke^{i\varphi'}$, from which it is clear that $ke^{i\varphi}+k'e^{i\varphi'}=2ke^{i(\varphi+\varphi')/2}\cos(\Delta\varphi/2)$, with $\Delta\varphi=\varphi-\varphi'$, which vanishes only for $\varphi'=\varphi+\pi\mod2\pi$. This provides a simple proof that back-scattering, $\bfk \rightarrow -\bfk$, is not allowed off a single point-like defect. At the same time, the expression shows that skew-scattering for any $\bfk'\neq-\bfk$ is generally allowed. This situation is illustrated in Fig. \ref{fig-Mperp} (a), which shows the magnetic response $M_\perp(\bfk,\bfk')=\|(M_x,M_y)(\bfk,\bfk')\|$, where $M_{x(y)}(\bfk,\bfk')=-\tr\im\, \sigma_{x(y)}\bfG(\bfk,\bfk')/2\pi$, for a fixed $\bfk=k(1,0)$ and energy $\omega=v_Fk$. It is clear that elastic skew-scattering off the single defect (dark ring-like feature) is present for all $\bfk'=ke^{i\phi'}$ with $\phi'\neq\pi$, although near-perfect back-scattering is notably suppressed.

%
%
Next, we consider $N$ point-like defects described by the Hamiltonian $\Hamil_\inter=\sum_{m=1}^N\Psi^\dagger(\bfr_m)\bfV_m\Psi(\bfr_m)$. with $\bfV_m=V_m\sigma^0\delta(\bfr-\bfr_m)$. Similarly to the above equation for a single defect, we here obtain the exact equation
\begin{subequations}
\label{eq-T2matrix}
\begin{align}
\bfG(\bfk,\bfk')=&
	\delta(\bfk-\bfk')\bfg(\bfk)
	+\sum_{mn}\bfg(\bfk)e^{-i\bfk\cdot\bfr_m}{\cal T}_{mn}e^{i\bfk'\cdot\bfr_n}\bfg(\bfk'),
\label{eq-GggT2G}
\\
{\cal T}_{mn}=&
	\bfV_m(\bft^{-1})_{mn}
\label{eq-T2first}
\\
\bft_{mn}=&
	\delta_{mn}\sigma^0-\bfg(\bfR_{mn})\bfV_n,
\label{eq-V2}
\end{align}
\end{subequations}
with $\bfR_{mn}=\bfr_m-\bfr_n$.
%
Here, we have also introduced the bare real space surface GF $\bfg(\bfr)=g_0(\bfr)\sigma^0+\bfg_1(\bfr)\cdot\bfsigma$, where $\bfg_1(\bfr)=g_\perp(\bfr)(\hat{\bfr}\times\hat{\bf e}_3)$, and
\begin{align}
g_0(\bfr)=&
	\frac{\omega}{i4v^2}H_0^{(1)}(\omega r/v),
&
g_\perp(\bfr)=&
	\frac{\omega}{4v^2}H_1^{(1)}(\omega r/v),
\label{eq-g0r}
\end{align}
whereas $H_m^{(1)}(x)$ is the Hankel function.
%
Considering an impurity comprising two point-like defects, $\bfV_m$, $m=1,2$, we can solve Eq. (\ref{eq-T2matrix}) analytically. Without loss of generality, we assume equal scattering potentials $V_m=V_0$. The correction $\delta\bfG(\bfk,\bfk')=\sum_{mn}g(\bfk)\exp(-i\bfk\cdot\bfr_n){\cal T}_{mn}\exp(i\bfk'\cdot\bfr_m)g(\bfk')$ can, after some lengthy algebra, be written as
\begin{align}
\delta\bfG(\bfk,\bfk')=&
	V_0^2
	\biggl\{
	\biggl(
		[V_0^{-1}-g_0]\Bigl[e^{-i(\bfk-\bfk')\cdot\bfr_1}+e^{-i(\bfk-\bfk')\cdot\bfr_2}\Bigr]
\nonumber\\&
		+g_0(\bfR_{12})\Bigl[e^{-i\bfk\cdot\bfr_1+i\bfk'\cdot\bfr_2}+e^{-i\bfk\cdot\bfr_2+i\bfk'\cdot\bfr_1}\Bigr]
	\biggr)
	\bfg(\bfk)\bfg(\bfk')
\nonumber\\&
		+\Bigl[e^{-i\bfk\cdot\bfr_1+i\bfk'\cdot\bfr_2}-e^{-i\bfk\cdot\bfr_2+i\bfk'\cdot\bfr_1}\Bigr]
		\bfg(\bfk)\bfg_1(\bfR_{12})\cdot\bfsigma\bfg(\bfk')
	\biggr\}
\end{align}
For the off-diagonal components of the correction we notice the following. The first contribution (factor in front of $\bfg(\bfk)\bfg(\bfk')$) vanishes for $\bfk'=-\bfk$ for the same reason as for the single defect. The second contribution (factor in front of $\bfg(\bfk)\bfg_1(\bfR_{12})\cdot\bfsigma\bfg(\bfk')$) can also be seen to vanish at $\bfk'=-\bfk$, since the difference $\exp(-i\bfk\cdot\bfr_1+i\bfk'\cdot\bfr_2)-\exp(-i\bfk\cdot\bfr_2+i\bfk'\cdot\bfr_1)\rightarrow0$ as $\bfk'\rightarrow-\bfk$. This difference describes two different scattering paths, which are related by time-reversal symmetry and exactly cancel each other at back-scattering. Summing up the two contributions shows that no channel for back-scattering opens from non-magnetic potential scattering, verifying previous lack of $180^\circ$ back-scattering results bases on Berry's phase arguments.\cite{Ando_98b}


Skew-scattering is however allowed for any $\bfk'\neq-\bfk$, which is clearly illustrated in Fig. \ref{fig-Mperp} (b), (c), where we plot the magnetic response $M_\perp(\bfk,\bfk')$ for two and three scattering points in the defect. A comparison between one, two, and three scattering points clearly shows both that skew-scattering is generally enhanced with the number of scattering points, and that near-perfect back-scattering becomes increasingly important.

\begin{figure}[t]
\begin{center}
\includegraphics[width=0.99\columnwidth]{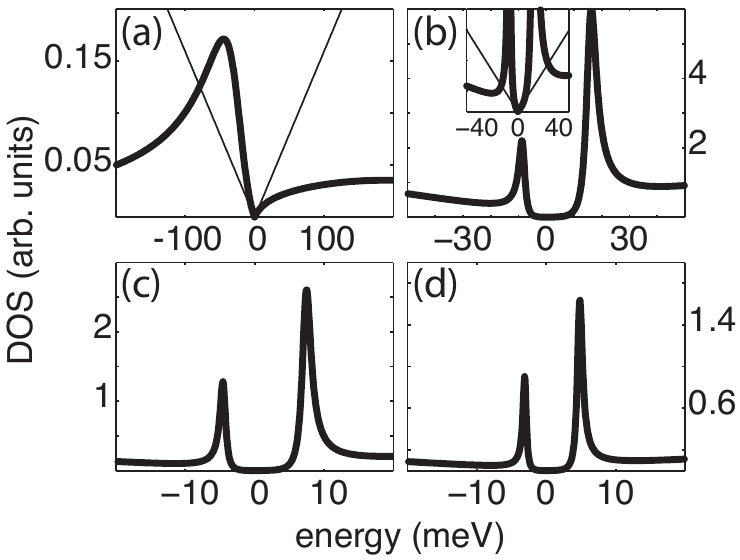}
\end{center}
\caption{Evolution of the local DOS for (a) one through (d) four scattering points located on a circle with 0.5~\AA\ radius. In panel (a, b), the DOS for pristine surface states (faint) is shown for reference. The inset in panel (b) shows the local DOS at a distance $\sqrt{2}$ \AA\ from the defect center, scaled up 60 times. Other parameters as in Fig. \ref{fig-Mperp}.
}
\label{fig-1}
\end{figure}

A very interesting question pertaining to our results is whether the enhanced skew-scattering leads to the emergence of a gapped DOS locally around the defects. It is a well-known theoretical result that purely magnetic scatterers generate a gap in the local DOS.\cite{SCZhangimp} The magnetic moment is effectively equivalent to a mass term in the Hamiltonian, acting as a coupling between the two linear bands which leads to a gap opening. As magnetic moments also allow spin-flip scattering, their presence enable prefect back-scattering. In the case of non-magnetic impurities there is no gap opening in the strict sense, but as we shall show below there is nonetheless a very clear gap-like feature emerging in the local DOS.

First, we consider the influence of the effective impurity potential on the spatial DOS. For any number of defects $N$, the $T$-matrix equation in real space is given by
\begin{align}
\bfG(\bfr,\bfr')=&
	\bfg(\bfr-\bfr')
	+\sum_{mn}\bfg(\bfr-\bfr_m){\cal T}_{mn}\bfg(\bfr_n-\bfr'),
\end{align}

In Fig.~\ref{fig-1} we plot the calculated LDOS for (a) one through (d) four scattering points. For the single defect (a), the LDOS is calculated at the defect site, showing a typical single impurity resonance below the Fermi level ($\dote{F}=0$) locally distorting the original Dirac dispersion (faint), in agreement with previous results.\cite{Biswas10, Black-Schaffer12imp,Black-Schaffer12imp2} In panels (b)-(d), the defects are evenly distributed along a circle with 0.5~\AA\ radius with the LDOS calculated in the center. The plots clearly show the appearance of a second peak above the Fermi level and the emergence of a gap-like feature inbetween the impurity resonances. The inset in panel (b) illustrates that the DOS between the resonances is finite and does not exhibit a gap in the strict sense. However, since the DOS is strikingly lower than the surrounding DOS we refer to this as a gap-like feature. For increasing number of scattering points the impurity resonances move closer to the Fermi level, similarly as found for increasing scattering potential $V_0$.\cite{Biswas10,Black-Schaffer12imp} Strong impurity resonances for non-magnetic impurities have also recently been observed in STS measurements.\cite{Teague12, Alpichshev_PRL12} This behavior reflects that the effective scattering potential experienced by the electrons increases with increasing number of scattering point located closely to one another. \cite{Hammar13}

In order to further analyze the physics emerging around the impurities we reduce our system to a single dimer of scattering points with equally strong scattering potentials $V_m=V_0$, $m=1,2$. We then have $\bft_{mm}=(1-g_0V_0)\sigma^0$, $m=1,2$, and $\bft_{mn}=-[g_0(\bfR_{mn})\sigma^0+\bfg_1(\bfR_{mn})\cdot\bfsigma]V_0$, $m\neq n$. Using these observations we can derive the following closed expression for the correction $\delta\bfG(\bfr,\bfr')$ to the real space GF
\begin{align}
\delta\bfG(\bfr,\bfr)=&
	2
	\frac{
		F_0(\bfr)+F_1(\bfr)+\bfF_2(\bfr)\cdot\bfsigma
	}
	{
		(1-g_0V_0)^2-V_0^2[g_0^2(\bfR_{12})-g_\perp^2(\bfR_{12})]
	}
	,
\end{align}
where $F_0(\bfr)=(1-g_0V_0)\sum_m[g_0^2(\bfR_m)-g_\perp^2(\bfR_m)]/2$, $F_1(\bfr)$, and $\bfF_2(\bfr)$ are functions of the electronic structure.\footnote{$F_1(\bfr)=g_0(\bfR_1)g_0(\bfR_{12})g_0(\bfR_2)
	-g_\perp(\bfR_1)g_0(\bfR_{12})g_\perp(\bfR_2)\hat\bfR_1\cdot\hat\bfR_2
	-\sum_{m\neq n}\sigma^z_{mm}g_0(\bfR_m)g_\perp(\bfR_{mn})g_\perp(\bfR_n)\hat\bfR_{mn}\cdot\hat\bfR_n
$, $\bfF_2(\bfr)=
	ig_\perp(\bfR_1)g_\perp(\bfR_{12})g_0(\bfR_2)\hat\bfR_1\times\hat\bfR_{12}
	+g_\perp(\bfR_1)g_\perp(\bfR_{12})g_\perp(\bfR_2)[\hat\bfR_1\times\hat\bfR_{12}]\times\hat\bfR_2$}

\begin{figure}[t]
\begin{center}
\includegraphics[width=0.99\columnwidth]{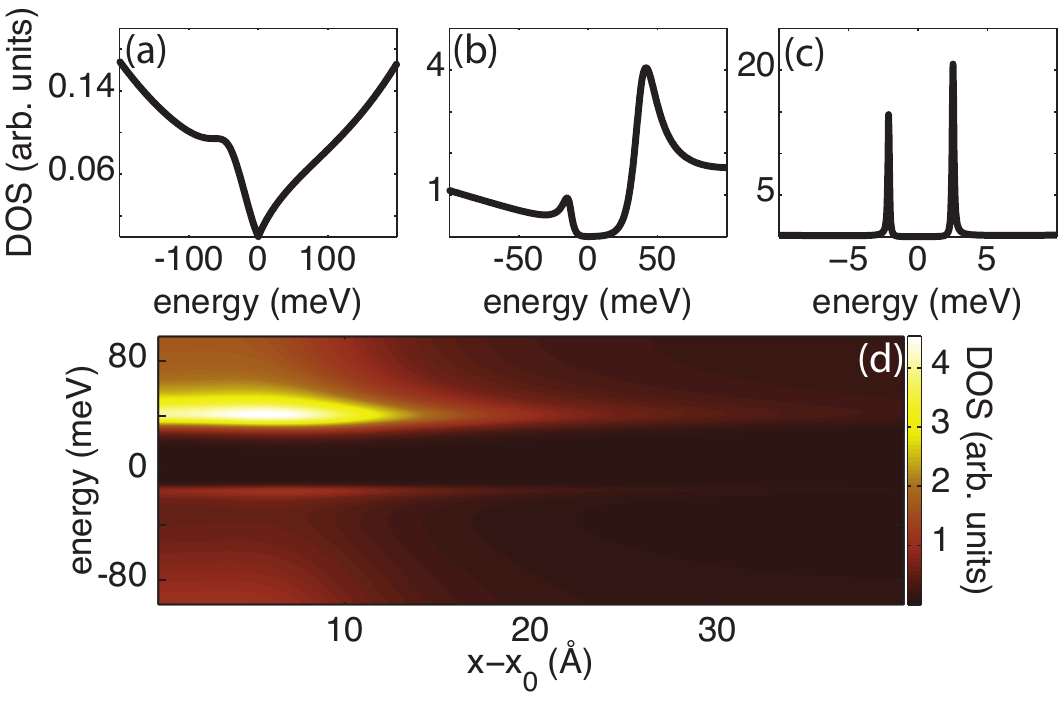}
\end{center}
\caption{Evolution of the local DOS for decreasing distance (a) through (c) $[10,\ 1,\ 0.1]$ \AA\ between two scattering points. Panel (d) shows the energy and spatial dependence of the DOS locally around the impurity dimer in panel (b). Other parameters are as in Fig. \ref{fig-1}.}
\label{fig-2}
\end{figure}

From this expression we can extract three pieces of important information. First, it is clear that only the the last contribution in the above expression is spin selective. However, as this contribution scales linearly for small spatial extensions $R_{12}\equiv|\bfR_{12}|$ of the impurity, its influence on the magnetic texture is negligible. Thus, the correction does not in any significant way alter the magnetic structure of the surface states.
Second, for large distances between the scattering points, $|\omega|R_{12}/v_F\gg1$, the second contribution in the denominator is negligible, such that the leading contribution effectively goes like $1/(1-g_0V_0)$. This property emphasizes the fact that the collection of scattering points should be regarded as a collective unit. Hence, the correction retains the character of a single defect,\cite{Biswas10} see Fig.~\ref{fig-2} (a). Third, for small distances between the scattering points, $|\omega|R_{12}/v_F\ll1$, there are two resonance peaks emerging at low energies, one at each side of the Fermi level. The correction contains the denominator $(1-g_0V_0)^2-V_0^2[g_0^2(\bfR_{12})-g_\perp^2(\bfR_{12})]$, and the factorization
\begin{align}
[1-g_0V_0-V_0{\cal R}e^{i\varphi/2}][1-g_0V_0+V_0{\cal R}e^{i\varphi/2}]
\end{align}
suggests that the first (second) factor represents an electron (hole) resonance, where ${\cal R}=\sqrt{|g_0^2(\bfR_{12})-g_\perp^2(\bfR_{12})|}$ and $\tan\varphi=\im[g_0^2(\bfR_{12})-g_\perp^2(\bfR_{12})]/\re[g_0^2(\bfR_{12})-g_\perp^2(\bfR_{12})]$. These resonances are conspicuous in Fig.~\ref{fig-2} (b) and (c), were we plot the local DOS for two scattering points for decreasing inter-impurity distances. For decreasing distance $|\bfR_{12}|$, the contribution $V_0{\cal R}\gg1$ indicates that the electron and hole resonances become more symmetrically located around the Fermi level. This behavior is verified in Fig.~\ref{fig-2} (b) and (c). The spatial extension of the emergent gap in the DOS locally around the impurity asymptotically scales as $\sqrt{v_F/|\omega|R_{12}}$. Thus, for $R_{12}\sim1$ \AA\ and $V_0=50$ meV the resonances emerge at about $10-50$ meV, which gives a spatial range of the order of 10-30 \AA, see Fig.~\ref{fig-2} (d).

\begin{figure}[t]
\begin{center}
\includegraphics[width=0.99\columnwidth]{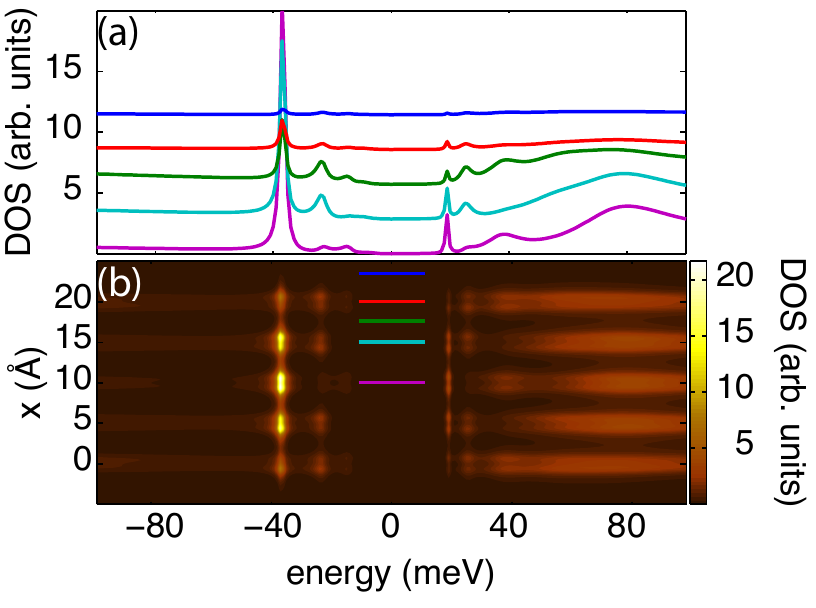}
\end{center}
\caption{(a,b) Energy and spatial dependence of the local DOS for five impurity dimers located at $x=0,5,10,15,20$ \AA. The distance between the scattering points within a dimer is set to 1 \AA. The traces in (a) correspond to the positions indicated by the lines in (b). Other parameters are as in Fig. \ref{fig-1}.}
\label{fig-3}
\end{figure}

The spatial dependence of the emergent gap can be exploited further to engineer structures with a spatially extended gap in the DOS. For example, distributing impurities in a linear chain on the surface creates a modified DOS along the structure with an apparent gap which is correspondingly extended in real space. In Fig.~\ref{fig-3} we plot the energy and spatial dependence of the LDOS near five impurity dimers. It is readily seen that the gap remains finite throughout the linear chain of impurities due to the finite spatial range of the modified DOS. The traces in panel (a) show the DOS at the positions indicated by lines in panel (b), more clearly illustrating the presence of the gap in an extended spatial range. These traces, moreover, clearly demonstrate the emergence of several resonance peaks on each side of the Fermi level. The appearance of more than just one electron and hole resonance is a result of multiple scatterings off the assembly of scattering points, as it would appear in a realistic structure.
The plots in Fig. \ref{fig-3} demonstrate that the structure of impurities create a local electronic structure which is markedly differ from the unperturbed linear spectrum. It is also clear that the electronic structure in the interior of the impurity chain is more developed than on the edges. By further extending the impurity chain to comprise even more impurity dimers, the interior electronic structure approaches a \emph{bulk} electronic structure, where a finite gap is generated on the surface of a topological insulator without breaking time-reversal symmetry.

We have, using a $T$-matrix approach, explicitly demonstrated that spatially extended impurities, here addressed through a collection of two or more non-magnetic scattering points, dramatically enhances both elastic skew-scattering and near-perfect $180^\circ$ back-scattering, showing that scattering is only suppressed at the singular perfect back-scattering point in $k$-space.
Since any realistic defect has a spatial extent and thus cannot be just modeled by a single-point defect, our results show that substantial skew- and near-perfect back-scattering is always present in real 3D TIs, even when time-reversal symmetry is intact. Further, multiple non-magnetic defects give rise to a local effective energy gap around the impurities, surrounded by sharp impurity-induced resonances. By engineering impurity structures, we have shown that it is possible to create an extended energy gap region on the surface of a 3D TI.


We are grateful to D. Abergel, S. Banerjee, J. Edge, Y. Kedem, S. Pershoguba, and K. Zakharchenko for stimulating and fruitful discussions. This work was supported by the Swedish Research Council, the G\"oran Gustafsson Foundation, the Knut and Alice Wallenberg Foundation, and the European Research Council under the European Union's Seventh Framework Program (FP/2207-2013)/ERC Grant Agreement DM-321031. Work at Los Alamos was supported by the US DoE Basic Sciences for the National Nuclear Security Administration of the US Department of Energy under Contract No. DE-AC52-06NA25396.

\bibliography{DiracMaterials}

\end{document}